\documentclass[a4paper]{article}
\usepackage{arxiv}
\usepackage{parskip}
\usepackage{pslatex}
\usepackage{hyperref}
\usepackage{graphicx}
\usepackage{amsmath}
\usepackage{amsfonts}
\usepackage{amssymb}
\usepackage{mathrsfs}
\usepackage[table]{xcolor}

\usepackage{graphicx}
\usepackage{amsmath}
\usepackage{amsfonts}
\usepackage{amssymb}

\title{Active Training of Physics-Informed Neural Networks to Aggregate and Interpolate Parametric Solutions to the Navier-Stokes Equations}

\author{ \href{https://orcid.org/0000-0002-0448-6146}{\includegraphics[scale=0.06]{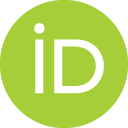}\hspace{1mm}Christopher J.~Arthurs}\thanks{Corresponding author.} \\
    School of Biomedical Engineering and Imaging Sciences \\
    King's College London \\
    4th Floor Lambeth Wing, St Thomas' Hospital, London SE1 7EH, UK\\
    \texttt{christopher.arthurs@kcl.ac.uk} \\
    \And
    \href{https://orcid.org/0000-0002-9965-7015}{\includegraphics[scale=0.06]{orcid.png}\hspace{1mm}Andrew P.~King} \\
    School of Biomedical Engineering and Imaging Sciences \\
    King's College London \\
    4th Floor Lambeth Wing, St Thomas' Hospital, London SE1 7EH, UK\\
    \texttt{andrew.king@kcl.ac.uk} \\
}

\begin{document}
\maketitle

\begin{abstract}The goal of this work is to train a neural network which approximates solutions to the Navier-Stokes equations across a region of parameter space, in which the parameters define physical properties such as domain shape and boundary conditions. The contributions of this work are threefold:
\begin{enumerate}
    \item To demonstrate that neural networks can be efficient aggregators of whole families of parameteric solutions to physical problems, trained using data created with traditional, trusted numerical methods such as finite elements. Advantages include extremely fast evaluation of pressure and velocity at any point in physical and parameter space (asymptotically, ~3 $\mu s$ / query), and data compression (the network requires 99\% less storage space compared to its own training data).
    \item To demonstrate that the neural networks can accurately interpolate between finite element solutions in parameter space, allowing them to be instantly queried for pressure and velocity field solutions to problems for which traditional simulations have never been performed.
    \item To introduce an active learning algorithm, so that during training, a finite element solver can automatically be queried to obtain additional training data in locations where the neural network's predictions are in most need of improvement, thus autonomously acquiring and efficiently distributing training data throughout parameter space.
\end{enumerate}
In addition to the obvious utility of Item 2, above, we demonstrate an application of the network in rapid parameter sweeping, very precisely predicting the degree of narrowing in a tube which would result in a 50\% increase in end-to-end pressure difference at a given flow rate. This capability could have applications in both medical diagnosis of arterial disease, and in computer-aided design.
\end{abstract}

\keywords{Navier-Stokes \and Deep Learning \and Active Learning \and Training \and Computing Methods}

\section{Introduction}
Many problems in physics are stated in terms of systems of partial differential equations (PDEs), whose solution fields represent some physical quantity of interest, such as fluid pressure and velocity, temperature, concentration, displacement, or electrical potential. These systems are routinely solved on high-performance computers using numerical solution schemes such as finite differences, finite volumes, or the finite element method (FEM). Such systems of equations can be very computationally expensive to resolve, and as such, it is desirable to perform as few simulations as possible, whilst extracting maximal scientific value from those that we do perform.

Recently, methods have been developed for encoding solutions to PDEs in artificial neural networks \cite{MR19}. The Universal Approximation Theorem (UAT) states that for any continuous function on a compact domain, there exists a neural network which approximates it arbitrarily well \cite{ZL17}. Solutions to PDEs fall into this category, so it is reasonable to expect that neural networks can be trained to provide surrogate solutions to PDEs. These networks are typically trained to encode the solution $u(x)$ to the PDE, so that it can be instantaneously queried for the solution at a location vector $x$. Previous work has focused on the power of the network to accurately represent $u$ at locations $x$ where training was performed, either by learning from classical FEM simulation results explicitly, or by directly inferring the solution via introducing a residual formulation of the PDE into the training loss function, together with a suitable encoding of the boundary conditions \cite{MR19}. These networks are referred to as physics-informed neural networks (PINNs). Such networks have found application throughout science and engineering, including fluid dynamics \cite{raissi2018hidden,MR19,rao2020physicsinformed}, material electromagnetic property discovery \cite{Chen:20}, acoustic waves \cite{YX19}, nonlinear diffusivity \cite{kadeethum2020physicsinformed}, material fatigue \cite{YA19}, and dynamical systems \cite{raissi2018multistep}. Very recently, initial investigations have been made into encoding solutions parametrically as $u(x, \theta)$, for $\theta$ some set of solution variables \cite{LS20}; $\theta$ may include domain shape, physical properties, or boundary condition parameters. This work builds on a body of previous efforts to infer parameters of - or solutions to - differential equations \cite{MR19,WP20,MR17b,MR18,MR18b}.

Building on our previous work \cite{CA20}, we introduce an active learning algorithm (ALA) for training PINNs to predict PDE solutions over entire regions $\mathcal{R}$ of parameter space, using training data from a minimal number of locations in $\mathcal{R}$. Active learning is a paradigm in which the network training procedure identifies and requests additional, high-benefit training data from an \textit{oracle} \cite{BS09,LLS10}. Typically, the oracle is a human, and the oracle's task is to label additional examples. This approach has been used in a number of scientific applications \cite{JS18,MR17a,DZ19}. In our case, because training data consists of classical PDE simulation solutions $u$ for physical parameter sets $\theta$, the oracle is a combination of a parametric finite element mesh generator and a FEM solver, and the ALA is fully autonomous. Previous works have explored active learning within single PDE models \cite{MR17a,DZ19}; the key contribution of this paper is to present a novel method for active learning across whole parametric families of models. To the best of our knowledge, this is the first work that integrates a learning algorithm, a domain and mesh generator, and a classical PDE solver so that the whole process bootstraps itself, and the algorithm is entirely autonomous.

This article is structured as follows. We present the active learning algorithm in detail, and use it to train a neural network to predict Navier-Stokes solutions $u(x, \theta)$ in a 2D domain with parametric shape and boundary conditions, for all $\theta \in \mathcal{R}$. We then evaluate the predictive accuracy using $\mathscr{L}^{2}$ norms of the difference between neural network and FEM solutions throughout the parameter space, focusing in particular on parameter locations where no training data was used. We then directly evaluate satisfaction of the PDE boundary conditions by the neural network predictions, by computing the $\ell^{2}$ error at a grid of points on the domain boundary. We demonstrate an application of the trained network to an inverse problem: predicting the tube shape parameter which will result in a 50\% increase in end-to-end pressure difference at a given flow rate, and then we discuss the advantages of neural networks in terms of both computational efficiency and data storage. In a number of places, we compare the ALA-trained network to a network trained instead using a random training data selection strategy.

\section{Methods}
\subsection{The Physical Problem}

\begin{figure*}[htbp]
    \centering
    \includegraphics[width=0.99\textwidth]{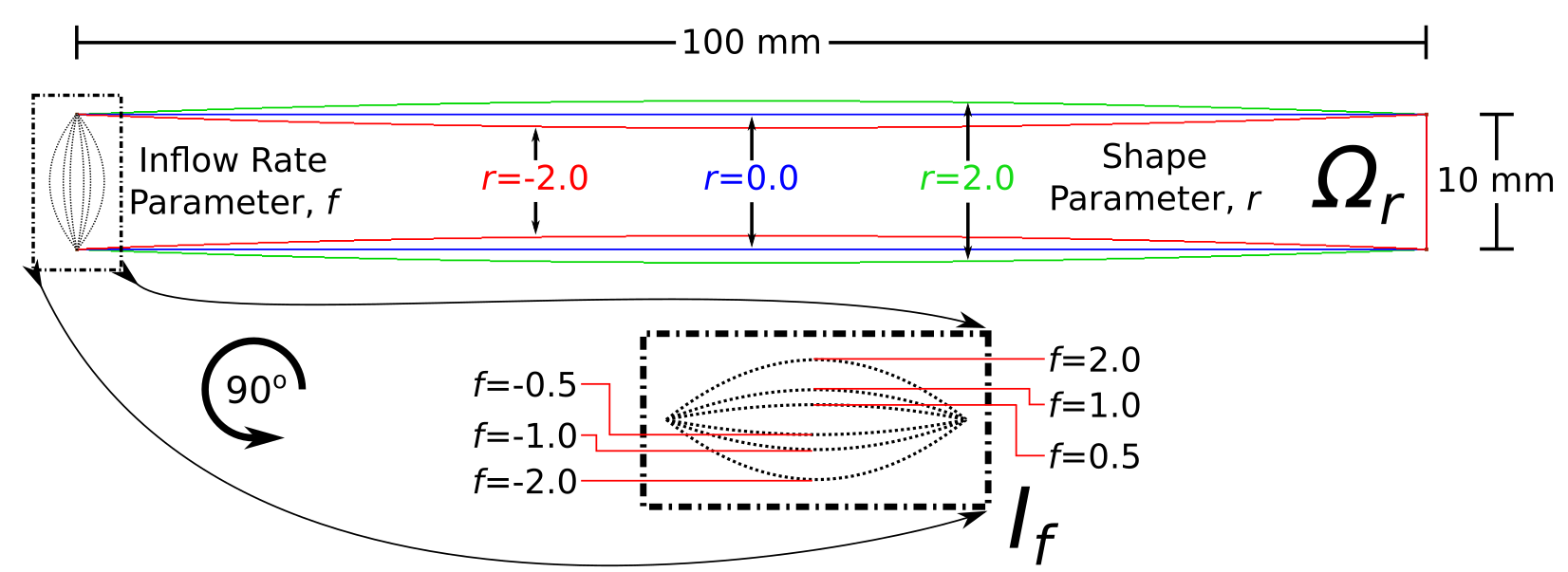}
    \caption{\small The parametric problem setup. The two parameters in question are the domain shape, $r$, and the inflow rate, $f$. We wish to train a neural network to predict pressure and velocity fields for values of parameters $r$ and $f$ which were not in the training data, and ultimately, everywhere in a region, $\mathcal{R}$, of parameter space. The domain is 100 $mm$ in length, and 10 $mm$ in diameter at the ends. Note that the inflow rate may be positive or negative, according to the sign of $f$.}
    \label{fig:parametric_domain}
\end{figure*}

We are interested in a parametric family of solutions to the steady incompressible Navier-Stokes equations in a continuous set of 2D tubular domains $\Omega_{r}$, with a continuous set of flow boundary conditions $I_{f}$, where $\theta=(f, r)$ parametrises the domain shape and the flow boundary condition; see Figure \ref{fig:parametric_domain}. $\Omega_{r}$ is $100~mm$ long, and at each end, $10~mm$ in width. The width towards the midpoint of the tube varies according to $r$, either stenosing or bulging outwards.

The Navier-Stokes equations, whose solutions throughout parameter space $\Omega_{r}$ we wish to encode into a neural network, are given by

\begin{eqnarray}
\rho \left(u \cdot \nabla\right) u - \mu \nabla^{2} u &=& -\nabla p, \label{eqn:navierstokes}\\
\nabla \cdot u &=& 0, \nonumber \\
\left(u, \frac{dp}{d\hat{n}}\right) &=& (0, 0) \mathrm{~on~} \partial \Omega_{r,w}, \nonumber \\
\left(u, \frac{dp}{d\hat{n}}\right) &=& (I_{f}, 0) \mathrm{~on~} \partial \Omega_{r,in}, \nonumber \\
\left(\frac{du}{d\hat{n}}, p\right) &=& (0, 0) \mathrm{~on~} \partial \Omega_{r,out}, \nonumber
\end{eqnarray}
where $\partial \Omega_{r,w}$ is the tube wall, $\partial\Omega_{r, in}$ is one end of the tube (``the inflow''), and $\partial\Omega_{r, out}$ the other (``the outflow''). $\hat{n}$ is an outward-pointing unit normal to the boundary. In the present work, independent of $r$, $\partial \Omega_{r,in} = \left\{(x,y) \in \mathbb{R}^{2} | x = 0;~ 0 \leq y \leq 10\right\}$, $\partial \Omega_{r,out} = \left\{(x,y) \in \mathbb{R}^{2} | x = 100;~ 0 \leq y \leq 10\right\}$, and we define $I_{f}(y):=f\cdot y\cdot(10-y)/25$. $u$ is the fluid velocity, and has units $mm\cdot s^{-1}$, and $p$ is the fluid pressure, with units $g \cdot mm^{-1} \cdot s^{-2} \equiv Pa$. The fluid density, $\rho=0.00106~g\cdot mm^{-3}$, and the dynamic viscosity $\mu=0.004~g\cdot mm^{-1}\cdot s^{-1}$; these values were chosen so that the problem is one of blood flow in a tube.

\subsection{Neural Network}
\begin{figure*}[htbp]
    \centering
    \includegraphics[width=1.0\textwidth]{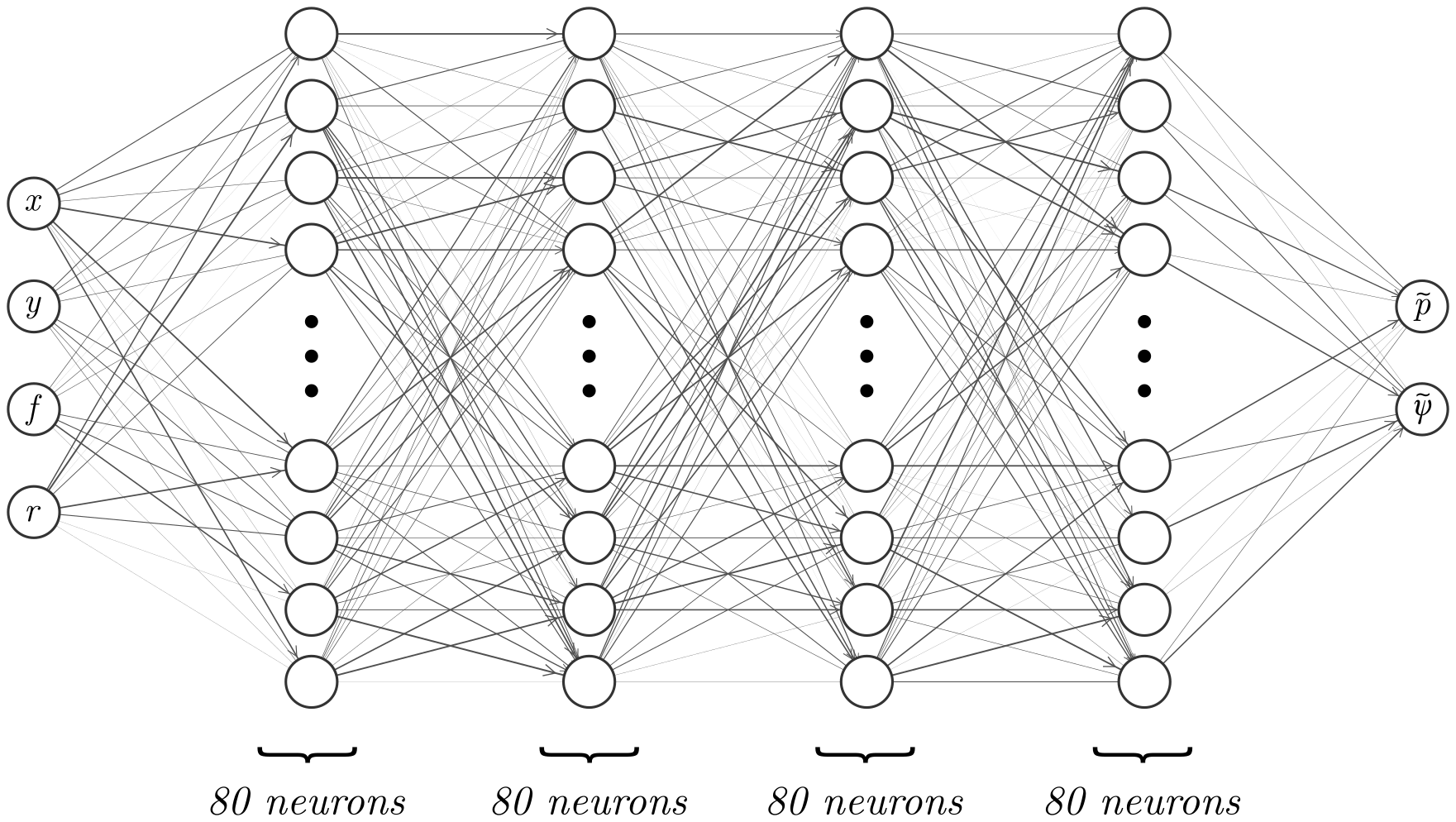}
    \caption{\small The neural network. A fully-connected network with four hidden layers and eighty neurons per hidden layer is used. There are four inputs: the two spatial coordinates, the peak inflow velocity $f$, and the tube shape parameter $r$. The two outputs are the pressure $\tilde{p}$, and a scalar potential for the velocity field $\tilde{\psi}$, from which the components of the velocity field can be determined by taking directional derivatives. Illustrative edge weights are indicated in greyscale. Figure created using open-source software \cite{AL19}.}
    \label{fig:neural_network}
\end{figure*}

The neural network that we wish to train is shown in Figure \ref{fig:neural_network}. It is fully-connected, uses hyperbolic tan activation functions, has four hidden layers with eighty neurons each, and has four scalar inputs and two scalar outputs. We require it to model the function $[\tilde{\psi}, \tilde{p}] = g(x, y, f, r)$, where $\tilde{\psi}$ is a scalar potential for the predicted velocity $\tilde{u}$, and $\tilde{p}$ is the pressure. Both are at location $(x, y)$ in physical space, and location $\theta=(r, f)$ in parameter space. Here, $r$ is the domain shape parameter, and $f$ is a peak Dirichlet boundary flow rate (see Figure \ref{fig:parametric_domain}). The fact that $\tilde{\psi}$ is a scalar potential for $\tilde{u}$ means that $\tilde{u}_{x} = \frac{d\tilde{\psi}}{dy}$ and $\tilde{u}_{y} = -\frac{d\tilde{\psi}}{dx}$, for the two-dimensional vector $\tilde{u}=[\tilde{u}_{x}, \tilde{u}_{y}]$. Throughout this work, we use tildes over variables to indicate that they are neural network predictions; otherwise they represent finite element solutions.

\subsection{Training Data}
The training data consist of velocity vectors $u_{i}=(u_{x}(x_{i}, y_{i}, \theta_{i}), u_{y}(x_{i}, y_{i}, \theta_{i}))$, for points indexed by $i \in \left[1, 2, 3, \dots, N\right]$. These data are generated by FEM simulation. The value of $N$ increases as the algorithm proceeds, as the ALA selects and generates additional training data, using an integrated mesh generator and FEM solver.

\subsection{Active Learning Algorithm, Network Training, Mesh Generation and the Finite Element Solver}
\label{sec:algorithm}
Core to the ALA is the definition of a region $\mathcal{R}$ in parameter space over which we want to train the neural network to produce accurate Navier-Stokes solutions. A grid, $\mathcal{G}$, of points, $\theta$, is placed over this region, creating a parameter discretisation with some choice of spacing. The purpose of the ALA is to iteratively identify points on this grid at which FEM data should be generated and added to the training set, in order to improve the quality of the neural network's predictions across the whole parameter region of interest. The goal is to train the network to accurately predict Navier-Stokes solutions everywhere in $\mathcal{R}$, using training data from as few of the points in $\mathcal{G}$ as possible. Our ALA falls into the category of pool-based selective sampling active learning algorithms \cite{BS09,LLS10}.

The active learning algorithm proceeds by the following steps.
\begin{enumerate}
\item (Initialisation) The initial training data is gathered, comprising $N$ points $(x_{i}, y_{i}, \theta_{i})$, at each of which $u$ is known from a finite element simulation. $N=a \times b \times c$, where $a$ is the number of velocity data points $u(x,y)$ per parameter point $\theta\in \mathcal{G}$, $b$ is the number of points $\theta\in\mathcal{G}$ at which we have FEM data, and $c$ is the proportion of the total data that we (randomly) select for use. In this work, $a\approx 1000$ (varying with domain shape), $b$ is initially 5 (as described in Section \ref{sec:ala_training}), and is incremented by one during Step 10 of this algorithm, and $c=1.0$.
\item (Training Step) The neural network is trained on the existing training data.
\item The network predicts pressure, $\tilde{p}$, and velocity, $\tilde{u}$, fields for all points in the parameter grid $\mathcal{G}$, regardless of where training data are available. The loss function term $L_{NS}$ is computed at all points of $\mathcal{G}$.
\item (Termination Condition) The ALA terminates if $L_{NS}$ is everywhere below some threshold value, or if training data are already available at all points of $\mathcal{G}$.
\item (Active Learning Step) The algorithm identifies a point on the parameter grid where the value of $L_{NS}$ is greatest.
\item A parametric description of the domain, using the parameters identified in Step 5, is automatically generated.
\item The domain is passed to the mesh generator, and a finite element simulation mesh is created.
\item The domain is pre-processed for finite element simulation, injecting the boundary condition parameters determined in Step 5.
\item The finite element solver runs the simulation.
\item The resulting velocity fields are appended to the training set, and $N$ is increased.
\item The algorithm returns to Step 2.
\end{enumerate}
We emphasise that the additional training data are appended to the existing training set, and the whole set is then used for further training, in order to avoid catastrophic forgetting.

During a Training Step, training takes place in two stages with two different optimisers. First, 20,000 iterations of the ADAM optimiser are performed, with a learning rate of 0.0001. Secondly, the network is passed to an L-BFGS-B optimiser to further refine the network weights, using a maximum of 50,000 iterations. The former is provided by Tensorflow \cite{tensorflow2015-whitepaper} and the latter by the Python package Scipy \cite{2020SciPy-NMeth}.

For 2D incompressible Navier-Stokes simulations, we use the Nektar++ finite element package \cite{CC15}. Simulations are performed on parametrically-defined 2D domains (Figure \ref{fig:parametric_domain}), generated programmatically according to the algorithmically-determined domain shape parameter $r$, using the Gmsh mesh generator \cite{CG09}.

\subsection{Loss Function}
The loss during training is computed according to
\begin{equation}
	L = \alpha_{u} L_{u} + \alpha_{NS} L_{NS} + \alpha_{p} L_{p} + \alpha_{BC} L_{BC},
	\label{eqn:lossfunction}
\end{equation}
where $L_{u}$ is the velocity loss, $L_{NS}$ is the Navier-Stokes residual loss, $L_{p}$ is a nodal reference pressure loss, and $L_{BC}$ is a boundary condition residual. Explicitly, with $\ell^{2}$ the standard Euclidean norm,
\begin{equation}
	L_{u} := \sum_{i=1}^{N} \left[\ell^{2}(u_{i} - \tilde{u}_{i})\right]^{2},
\end{equation}
for $\tilde{u}_{i}=\tilde{u}(x_{i}, y_{i}, \theta_{i})$ the neural network's prediction of the velocity at the $i$-th training point;
\begin{equation}
	L_{NS} := \sum_{i=1}^{N}\left[\ell^{2}\left(\rho \left(\tilde{u}_{i} \cdot \nabla\right) \tilde{u}_{i} + \nabla{\tilde{p}_{i}} - \mu \nabla^{2} \tilde{u}_{i}\right)\right]^{2} \label{eqn:lossfunction_navier_stokes},
\end{equation}
where $\tilde{p}_{i}=\tilde{p}(x_{i}, y_{i}, \theta_{i})$ is the neural network's prediction of the pressure at the $i$-th training point (cf. Equation \ref{eqn:navierstokes}). Note in particular that no training data appear in $L_{NS}$.
\begin{equation}
	L_{p} := \sum_{j=1}^{M}\left[\ell^{2} \left(\hat{p}_{j} - \tilde{\hat{p}}_{j}\right)\right]^{2} \label{eqn:lossfunction_pressure},
\end{equation}
for training data pressures $\hat{p}_{j}=p(\hat{x}, \hat{y}, \theta_{j})$, with $\theta_{j},~ j \in \left[1, 2, \dots, M\right]$, the $M$ points in $\mathcal{G}$ for which training data are available. The point $(\hat{x}, \hat{y})$ is some fixed location in the spatial domain, and the $\tilde{\hat{p}}_{j}=\tilde{p}(\hat{x}, \hat{y}, \theta_{j})$ are the neural network's corresponding pressure predictions. $L_{p}$ ensures that $\tilde{p}$ is fully defined, as opposed to defined up to a constant; thus it is analogous to having a Dirichlet boundary condition in Laplace's Equation.

Finally, satisfaction of the velocity boundary conditions is enforced by
\begin{equation}
	L_{BC} := \sum_{j=1}^{M} \sum_{k=1}^{K_{j}}\left[\ell^{2}\left(\tilde{u}_{j,k}-g_{j,k}\right)\right]^{2}, \label{eqn:bc_loss}
\end{equation}
for $\tilde{u}_{j,k}=\tilde{u}(x_{j,k}, y_{j,k}, \theta_{j})$ the neural network's predicted solution at $(x_{j,k}, y_{j,k}) \in \partial \Omega_{D,j}$, where $k\in[1,2,\dots,K_{j}]$ indexes all $K_{j}$ points in the training data which lie on $\partial\Omega_{D,j} := \partial\Omega_{r_{j}, w} \cup \partial\Omega_{r_{j}, in} $, for parameters $\theta_{j}$, $j \in [1, 2,\dots, M]$, and $K_{j}$ is $j$-dependent because there may be different numbers of training data points on $\partial\Omega_{D,j}$ for different domain shapes, as determined by $\theta_{j}=(f_{j}, r_{j})$. In Equation \ref{eqn:bc_loss},
\begin{equation}
	g_{j,k}=u(x_{j,k}, y_{j,k}, \theta_{j}):= 
	\begin{cases}
	0 & \mathrm{if~}(x_{j,k}, y_{j,k}) \in \partial \Omega_{r_{j},w} \\
	I_{f_{j}}(y_{j,k}) & \mathrm{if~}(x_{j,k}, y_{j,k}) \in \partial \Omega_{r_{j},in}
	\end{cases}
\end{equation}
is the imposed boundary condition; cf. Equation \ref{eqn:navierstokes}.

The parameters $\alpha$ in Equation \ref{eqn:lossfunction} give the relative importance of each component of $L$. During minimisation of $L$, a larger $\alpha$ for a component will result in more aggressive optimisation of that component. We found empirically that $\alpha_{u}=1,$ $\alpha_{NS}=10^{6},$ $\alpha_{p}=10^2$ and $\alpha_{BC}=10^{6}$ give rapid convergence to the accurate solutions which we desire.

Note that in the present work, the ALA chooses new points of $\mathcal{G}$ for the training set using only $L_{NS}$. Other strategies are likely equally valid.

\section{Results}
\subsection{Training the Network on a Parameter Space Region, $\mathcal{R}$}
\label{sec:ala_training}
We begin by training the neural network to produce Navier-Stokes solutions for all parameters $(f,r) \in \mathcal{R}=[-2, 2]^{2}\subset \mathbb{R}^{2}$. We discretise this with a grid of potential training points \newline $\mathcal{G}:=\left\{(f,r) \in \mathbb{R}^{2} ~|~ f,~r\in\left\{0, \pm \frac{1}{3}, \pm \frac{2}{3}, \dots, \pm \frac{6}{3}\right\}\right\}$. The network weights for the $N_{in}$ input and $N_{out}$ output connections of a neuron $\mathcal{N}$ were initialised randomly from a truncated normal distribution, with mean zero and variance $2/(N_{in}+N_{out})$ \cite{XG10}. Initial training data consisted of the Navier-Stokes FEM solution at the corners and centre of $\mathcal{R}$; i.e. all combinations of $(f,r)=(\pm 2.0, \pm2.0)$, and $(f,r)=(0.0, 0.0)$ - five points in total. We refer to this as \textit{corners-and-centre} initialisation. The algorithm in Section \ref{sec:algorithm} was run; 42 ALA iterations were completed, meaning that the algorithm identified and generated FEM training data at 42 points in $\mathcal{G}$.

Figure \ref{fig:lossplots} shows how the loss over $\mathcal{G}$ evolves as more ALA iterations are completed. We refer to these as \textit{active learning plots}, because they show how the ALA selects each additional training point. Red dots indicate locations where training data were generated and utilised. Note that after earlier iterations, the loss is low only where training data were provided, whereas later on, the loss is low even where no data were available. These plots make it clear that the quality of the solution improves as the ALA uses the FEM solver to add the most appropriate training data.

\subsection{Alternative Training Strategy - Random Data Selection}
\begin{figure*}[htbp]
    \centering
    \includegraphics[width=0.85\textwidth]{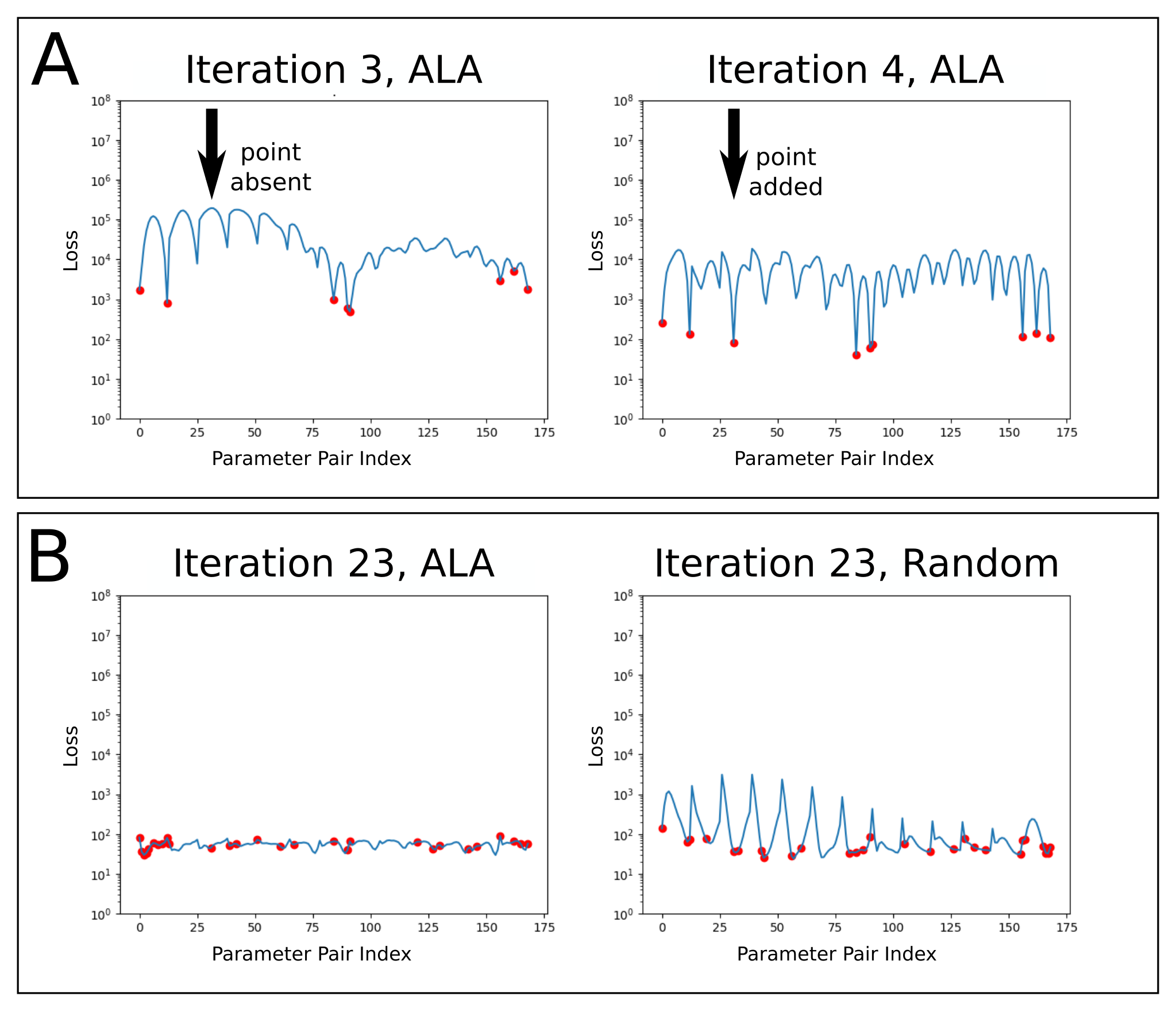}
    \caption{\small \textbf{Panel A:} Example of $L_{NS}$ during training, before and after an ALA iteration, showing the addition of FEM training data at one point in $\mathcal{G}$. The x-axis gives a row-major indexing of $\mathcal{G}$, and the y-axis gives $L_{NS}$ at each point. Red dots indicate points at which FEM training data was available. Note that training data was added at the point of maximum loss. \textbf{Panel B:} A comparison between the ALA training data selection strategy, and the random data selection strategy. $L_{NS}$ is shown across $\mathcal{G}$ after 23 iterations in both cases. Globally, the loss is greatly reduced in the ALA case.}
    \label{fig:lossplots}
\end{figure*}

To provide a point of comparison, we separately train a network using a random training data selection strategy, initialised in the same way as described in Section \ref{sec:ala_training}. In this case, the algorithm of section \ref{sec:algorithm} is followed, but Step 4 is replaced with random selection of an unused point of $\mathcal{G}$. Panel B of Figure \ref{fig:lossplots} shows a comparison between the ALA and the random strategy after 23 iterations of each. It is clear that the error, as indicated by $L_{NS}$, is far smaller and far more consistent under the ALA.

\subsection{Pressure and Velocity Field Errors in the $\mathscr{L}^{2}$ Norm}
The loss function $L$ is used by the ALA because its evaluation does not require the availability of a ground-truth FEM solution. This is key, because the ALA aims to require as few FEM solutions as possible to obtain accurate solutions everywhere in $\mathcal{R}$. However, the gold standard for evaluating solution accuracy is to compare with a ground-truth solution in the $\mathscr{L}^{2}$ norm, given by

\begin{equation}
	\left\| u - \tilde{u} \right\|_{2}=\sqrt{\int_{\Omega_{r}} (u - \tilde{u})^{2}~dA},
\end{equation}
where $u$ and $\tilde{u}$ are the finite element and neural network predictions of the velocity field, respectively. The $\mathscr{L}^{2}$ error for the pressure field is computed analogously. Because this is an integral with respect to the area measure $dA$, $u$ and $\tilde{u}$ must both be represented here using linear basis functions on the finite element mesh on which $u$ was computed. This error should not be routinely computed during applications of the ALA, because it defeats the purpose of minimising the number of FEM simulations; however, it allows us to validate the efficacy of the ALA.

\begin{figure*}[htbp]
    \centering
    \includegraphics[width=1.0\textwidth]{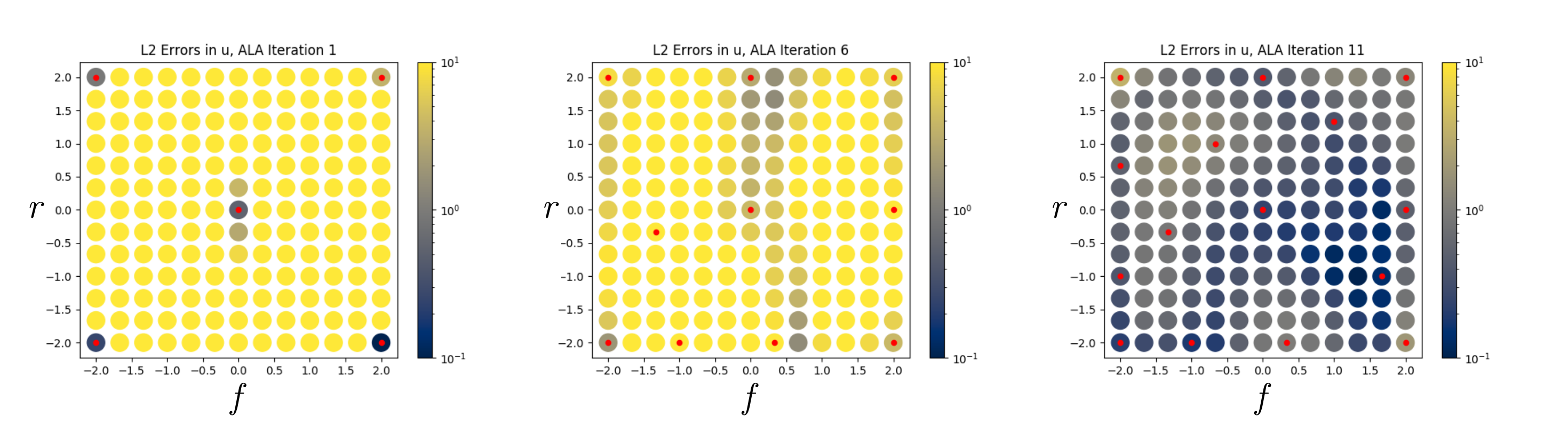}
    \caption{\small $L^{2}$ errors in the neural network's prediction of the velocity field, $\tilde{u}$, compared to the FEM solution, across parameter space $\mathcal{G}$, as the ALA iterations progress. Red dots indicate locations where training data were used.}
    \label{fig:l2_u}
\end{figure*}

\begin{figure*}[htbp]
    \centering
    \includegraphics[width=1.0\textwidth]{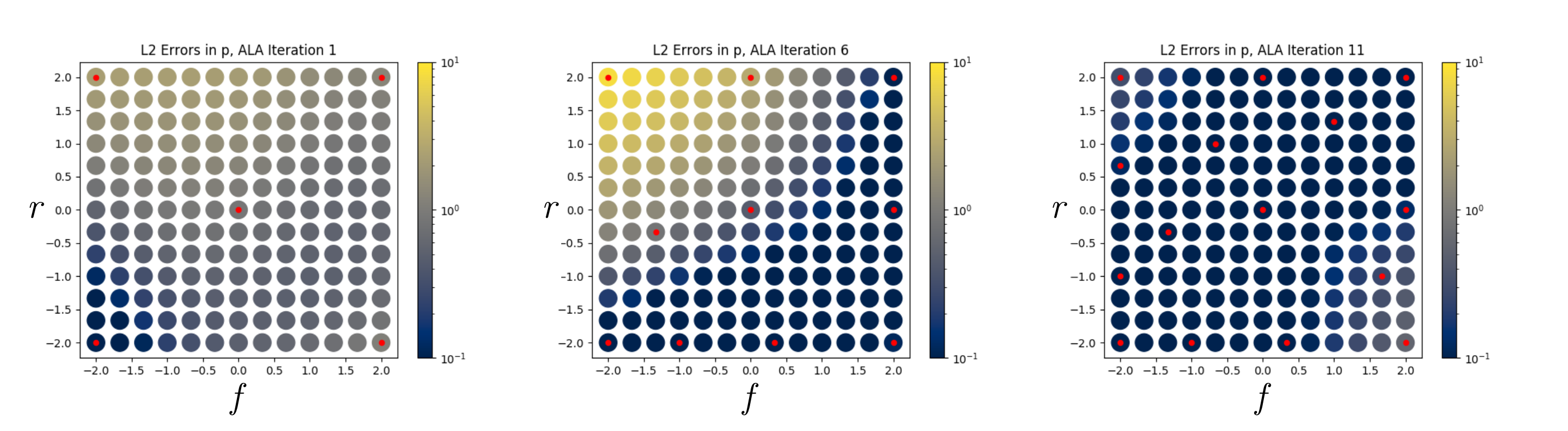}
    \caption{\small $L^{2}$ errors in the neural network's prediction of the pressure field, $\tilde{p}$, compared to the FEM solution, across parameter space $\mathcal{G}$, as the ALA iterations progress. Red dots indicate locations where training data were used.}
    \label{fig:l2_p}
\end{figure*}

Figures \ref{fig:l2_u} and \ref{fig:l2_p} show the $\mathscr{L}^{2}$ error, in the velocity and pressure fields respectively, between the neural network's prediction and the FEM solution, at all points of $\mathcal{G}$. Both figures show three plots, each separated by five iterations of the ALA. Red dots indicate the location of FEM training data, and the $\mathscr{L}^{2}$ error is shown by the colour scale. In Figure \ref{fig:l2_u}, we observe that the velocity field error is initially only low at locations where training data are available, but regions of parameter space with lower error rapidly start to form. The locally non-monotonic convergence of error is also evident at the corners. In Figure \ref{fig:l2_p}, we observe that the error in the pressure field rapidly becaomes small globally. This demonstrates that neural networks can aggregate and interpolate parametric solutions to the Navier-Stokes equations, as desired.

\subsection{Examples of Pressure and Velocity Field Predictions}
\begin{figure*}[htbp]
    \centering
    \includegraphics[width=0.85\textwidth]{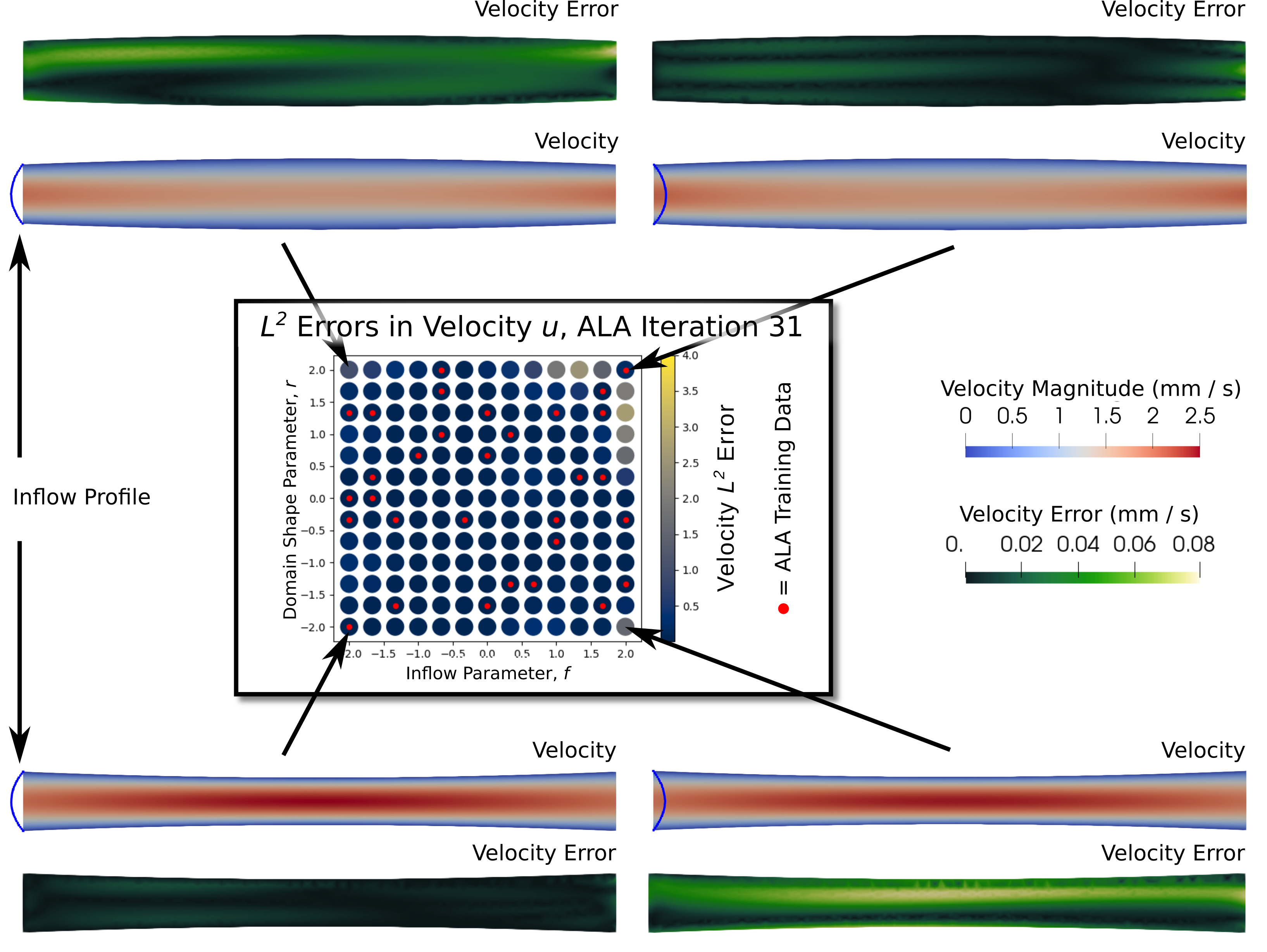}
    \caption{\small \textbf{Centre:} the $\mathscr{L}^{2}$ errors in the velocity field at all points in $\mathcal{G}$ after 31 ALA iterations, using a network which was initialised with training data only at $(r, f)=(2.0, 2.0)$, as opposed to the corners-and-centre initialisation used elsewhere in this work. Red dots indicate locations where training data were available. \textbf{Corners:} velocity predictions and local $\ell^{2}$ velocity errors after 31 ALA iterations, at four points in $\mathcal{G}$ representing the extremes of the training region, $\mathcal{R}$. The inflow profiles, whose peak is controlled by the parameter $f$, are shown at the left hand boundary of each tube, where they are imposed as boundary conditions. Errors are computed against the FEM ground-truth. Note that only two of the four predictions shown had training data by this stage of the ALA.}
    \label{fig:tube_plots_step_31}
\end{figure*}

In this section, we examine predictions on a neural network which was initially trained on data in only the top-right corner of $\mathcal{R}$, $(2.0, 2.0)$, as opposed to the corners-and-centre initialisation for other results in the paper. This allows us to compare predictions at corners with vs. corners without training data. We note that in general, corner-and-centre initialisation results in fewer ALA iterations being required before the network is globally accurate. Figure \ref{fig:tube_plots_step_31} shows velocity predictions from a neural network after 31 ALA iterations, together with the parameter-local $\mathscr{L}^{2}$ errors (central grid), and spatially-local $\ell^{2}$ errors in those predictions (green tube plots), using FEM as the ground-truth. Note that only two of the four examples had training data, but both $\mathscr{L}^{2}$ and $\ell^{2}$ errors in the velocity field are small in all cases. Because in the cases shown, $f=\pm 2.0$, by symmetry we expect the same velocity magnitude fields in the top two and the bottom two tubes shown.

Whilst the velocity errors are all very small - less than than 80 $\mu~m\cdot s^{-1}$ everywhere, and generally much smaller - they are towards the high end of this range in the top-left and bottom-right corners of $\mathcal{G}$ (Figure \ref{fig:tube_plots_step_31}). There are two reasons for this. The first is that we expect lower accuracy where no training data were used. The second is that these points are outside the convex hull of the training data in parameter space, and so the predictions are clearly extrapolatory. In general, networks should not be used to extrapolate, despite the fact that the extrapolatory solutions shown are actually quite accurate.

\subsection{Boundary Condition Errors in the $\ell^{2}$ Norm}
\begin{figure*}[htbp]
    \centering
    \includegraphics[width=1.0\textwidth]{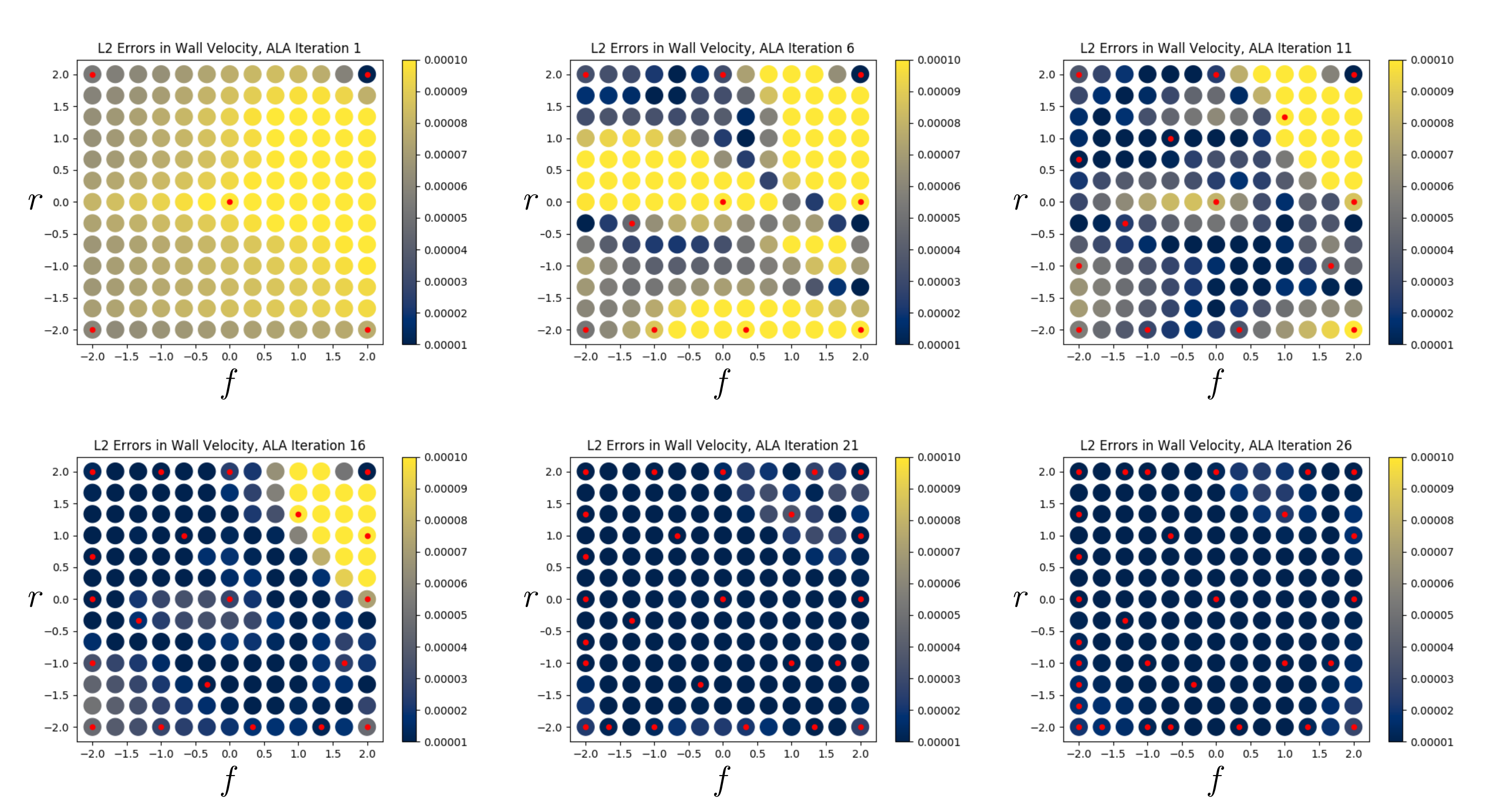}
    \caption{\small $\ell^{2}$ finite element nodal errors in the velocity field on the wall, across parameter space $\mathcal{G}$, as the ALA iterations progress. Red dots indicate locations where training data were used.}
    \label{fig:l2_noslip_velocity}
\end{figure*}

\begin{figure*}[htbp]
    \centering
    \includegraphics[width=1.0\textwidth]{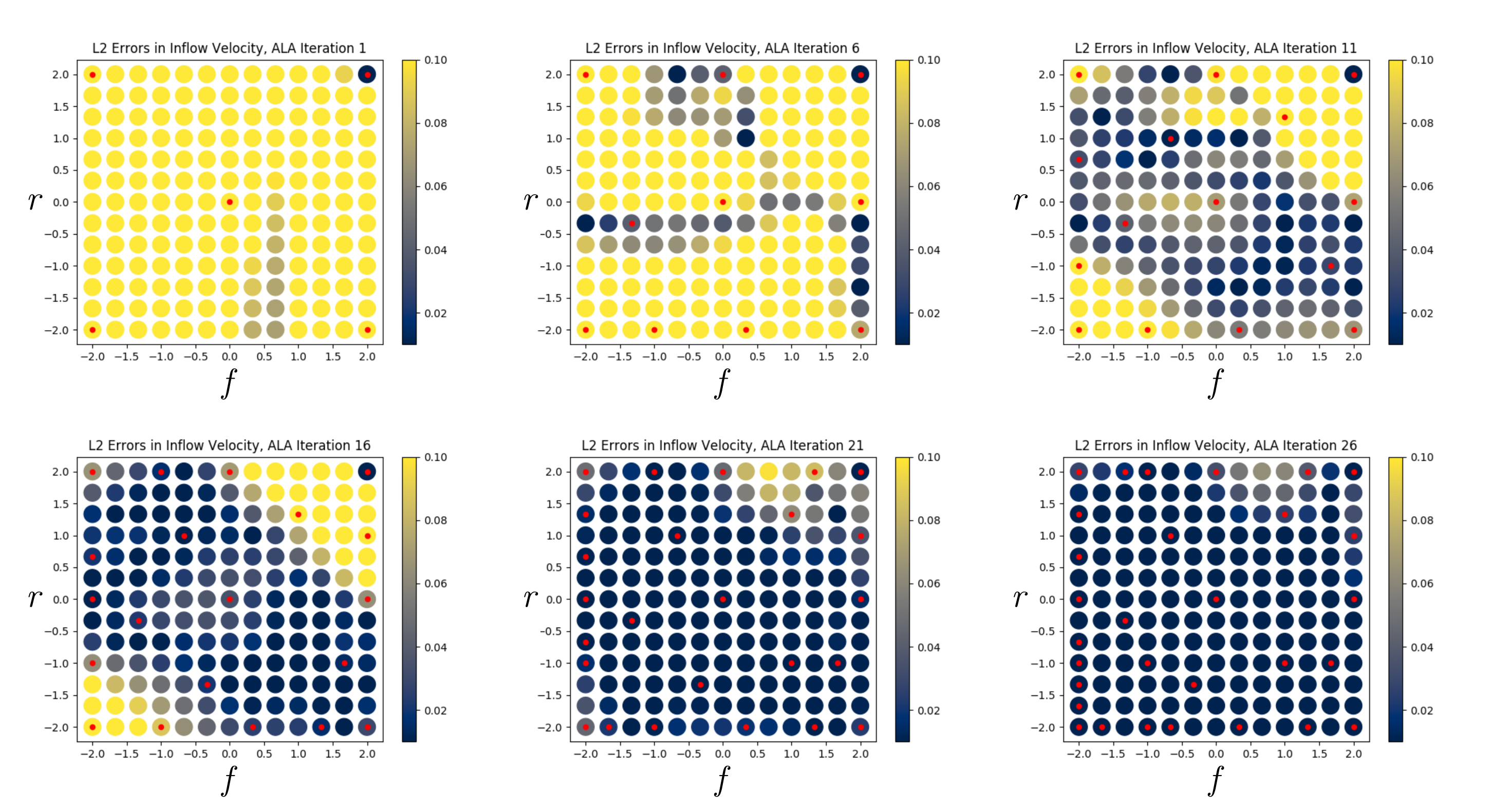}
    \caption{\small $\ell^{2}$ finite element nodal errors in the velocity field at the inflow boundary, across parameter space $\mathcal{G}$, as the ALA iterations progress. Red dots indicate locations where training data were used.}
    \label{fig:l2_inflow_velocity}
\end{figure*}

It is important that the boundary conditions are satisfied by the neural network's predicted solutions, whether or not training data were used for a particular parameter set. Figure \ref{fig:l2_noslip_velocity} shows the mean nodal $\ell^{2}$ error in the velocity field at the tube walls (i.e. the mean square root of Equation \ref{eqn:bc_loss}, but restricted to $\Omega_{r,w}$, and to one value of $j$ for each point in the figure), and Figure \ref{fig:l2_inflow_velocity} shows the mean nodal error in the velocity field at the inflow boundary (i.e. the mean square-root Equation \ref{eqn:bc_loss}, restricted to $\Omega_{r,in}$, and to one value of $j$ for each point in the figure). Thus, both figures break down Equation \ref{eqn:bc_loss} by parameter $\theta$ (i.e. by index $j$), rather than summing over them. We see that in both cases, the error at the boundary becomes small across all of $\mathcal{G}$ as the ALA progresses. This provides confidence that the loss function (Equation \ref{eqn:lossfunction}), and in particular its constituent $L_{BC}$, successfully ensures the boundary conditions are satisfied.

\subsection{Finding Model Parameters for Specific Physical Properties using the Neural Network}
\label{sec:parameter_sweep}

We now demonstrate the use of the neural network to rapidly search $\mathcal{R}$ for a tube design with a particular, arbitrary physical property.

\subsubsection{Finding the Tube Shape Parameter, $r$, which gives a 50\% Pressure Loss when $f=1.5$}
\label{sec:parameter_sweep_1pt5}
\begin{figure*}[htbp]
    \centering
    \includegraphics[width=1.0\textwidth]{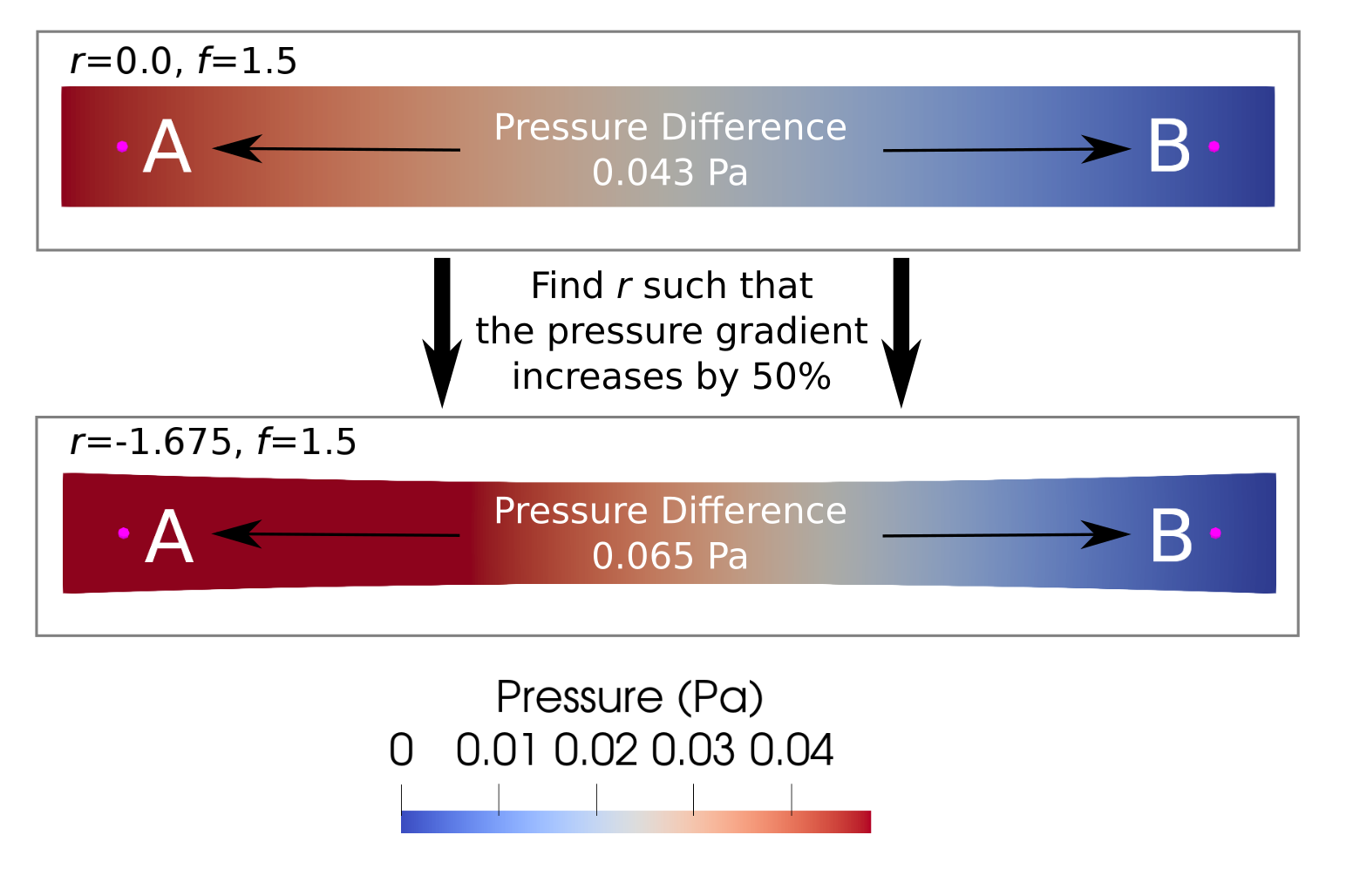}
    \caption{\small The pressure field in a tube with two different shape parameters $r$, computed using FEM. The upper panel shows a straight tube, and the lower panel a narrowing tube ($r=-1.675$) with a pressure difference between points $A$ and $B$ which is 50\% larger than in the straight tube. The parameter $r=-1.675$ was predicted as giving this 50\% increase in pressure difference by the neural network after 31 ALA iterations. This figure shows the FEM confirmation of the accuracy of that prediction. Cf. Table \ref{tab:f1pt5_50pct_narrowing_study}.}
    \label{fig:AB_pressure_difference}
\end{figure*}

In a straight, unobstructed tube ($r=0.0$), with an inflow parameter $f=1.5$, the FEM pressure difference $\Delta p$ between points A and B (upper panel of Figure \ref{fig:AB_pressure_difference}) is $0.043~Pa$. We wish to determine the degree of tube narrowing (parameter $r$) which causes $\Delta p$ to increase by 50\% over the this value, i.e. to $\Delta p = 0.065~Pa$ (lower panel of Figure \ref{fig:AB_pressure_difference}), at the same inflow parameter $f=1.5$.

We examine the neural network's predictions at two stages for which $L_{NS}$ is less than $100$ everywhere on $\mathcal{G}$ - specifically, after 23 and 34 ALA iterations\footnote{The maximum global loss does not decrease monotonically with ALA iterations; rather, it displays a general decreasing trend.}. We query these at all values of $r$ between $0$ and $-2.0$, using a step-size of $0.025$ (i.e. 81 different tube geometries). The results, together with the FEM simulation ground-truth, are shown in Table \ref{tab:f1pt5_50pct_narrowing_study}. Cells representing the 50\% increase are highlighted in green. We observe that the neural network prediction agrees perfectly in this metric with the FEM solutions: $\Delta p = 0.065~Pa$ occurs at both $r=-1.675$ and $r=-1.700$. For comparison, we also show the predictions of the neural network with 23 and 34 random parameter points included in the training data; we observe that the predictions in this case are of much lower quality, with errors in predicted $r$ ranging from $5.8\%$ to $10.3\%$, contrasting starkly with the zero error in the ALA predictions. This error is unsurprising, as at no point during the 42 training iterations using random data selection did we observe $L_{NS}$ dropping globally below $1000$ on $\mathcal{G}$. This demonstrates the efficacy of ALA training, its superiority to random training, the interpolative power of an ALA trained network, and its application to inverse problems.

The pressure field when $r=-1.675$ is shown in the lower panel of Figure \ref{fig:AB_pressure_difference}. The parameter sweep using the neural network took 7.6 seconds, of which around 3 $\mu s$ were required per value of $r$, with essentially all the remaining run-time being due to loading the trained network and transferring it to the GPU. Performing the same sweep with FEM would take 54 minutes - i.e. over 400 times longer. This is discussed further in Section \ref{sec:computational_cost}.

\begin{table}[h!]
\caption{Pressure differences $\Delta p$ (Pa) between points A and B in the tube shown in Figure \ref{fig:AB_pressure_difference}. A fixed value of $f=1.5$ is used, and we test the neural network after 23 (ALA (23)) and 34 (ALA (34)) iterations, parameter-sweeping over $r$, in search of a value which increases $\Delta p$ by 50\% over the straight tube ($r=0$), to 0.065 Pa (shown in green). FEM-derived values of $\Delta p$ are provided as a ground-truth. We include also the results of a neural network trained using a random data addition strategy, for each of 23 (Random (23)) and 34 (Random (34)) data-addition iterations. This table is discussed in Section \ref{sec:parameter_sweep_1pt5}.}
\label{tab:f1pt5_50pct_narrowing_study}
\begin{center}
\begin{tabular}{*{6}{c}}
\hline
$r$      &   FEM                         & ALA (23)                   & ALA (34)                     &  Random (23)                 &  Random (34)                \\
\hline
-1.650   &   0.064                       & 0.064                      &  0.064                       &  0.062                       &  0.061                      \\
\hline
-1.675   &   \cellcolor{green!25} 0.065  & \cellcolor{green!25} 0.065 &  \cellcolor{green!25} 0.065  &  0.063                       &  0.062                      \\
\hline
-1.700   &   \cellcolor{green!25} 0.065  & \cellcolor{green!25} 0.065 &  \cellcolor{green!25} 0.065  &  0.063                       &  0.062                      \\
\hline
-1.725   &   0.066                       & 0.066                      &  0.066                       &  0.063                       &  0.063                      \\
\hline
-1.750   &   0.066                       & 0.066                      &  0.066                       &  0.064                       &  0.063                      \\
\hline
-1.775   &   Not Run                     & 0.066                      &  0.066                       &  0.064                       &  0.063                      \\
\hline
-1.800   &   Not Run                     & 0.067                      &  0.067                       &  \cellcolor{green!25} 0.065  &  0.064                      \\
\hline
-1.825   &   Not Run                     & 0.067                      &  0.067                       &  \cellcolor{green!25} 0.065  &  0.064                      \\
\hline
-1.850   &   Not Run                     & 0.067                      &  0.068                       &  \cellcolor{green!25} 0.065  &  \cellcolor{green!25} 0.065 \\
\hline
-1.875   &   Not Run                     & 0.068                      &  0.068                       &  0.066                       &  \cellcolor{green!25} 0.065 \\
\hline
-1.900   &   Not Run                     & 0.068                      &  0.068                       &  0.066                       &  0.066                      \\
\hline
\end{tabular}
\end{center}
\end{table}

\subsubsection{Finding the Tube Shape Parameter, $r$, which Increases Pressure Loss by 50\% when $f=2.0$}
\label{sec:parameter_sweep_2pt0}
Performing the same experiment as described in Section \ref{sec:parameter_sweep_1pt5}, but this time using $f=2.0$, the FEM pressure difference in a straight tube ($r=0.0$) between points A and B (indicated in Figure \ref{fig:AB_pressure_difference}) is $\Delta p = 0.058~Pa$. We wish to determine $r$ such that $\Delta p$ increases by 50\%, to $\Delta p = 0.087~Pa$.

Querying the neural network at 81 distinct values of $r$ (every $0.025$ between $r=-2.0$ and $r=0.0$, inclusive), we obtain predictions $\Delta p = 0.086~Pa$ at $r=-1.700$, and $\Delta p = 0.087~Pa$ at the subsequent value, $r=-1.725$. These results are highlighted in green in Table \ref{tab:f2pt0_50pct_narrowing_study}. We observe that the agreement with FEM simulation is good, but not as perfect as we observed in the $f=1.5$ case. This is discussed further in Section \ref{sec:interpolation_vs_extrapolation}, but here we note that no further improvement in accuracy on this question was observed with further ALA iterations.

\begin{table}[h!]
\caption{Pressure differences $\Delta p$ (Pa) between points A and B in the tube (c.f. Figure \ref{fig:AB_pressure_difference}). With the inflow parameter $f=2.0$, we look for a value of $r$ which gives an increase in $\Delta p$ of 50\% over the value in the straight tube, to the target value of 0.87 Pa (highlighted in green). FEM-derived ground truth is compared with the neural network trained for 23 (ALA (23)), and 34 (ALA(34)) iterations. This table is discussed in Sections \ref{sec:interpolation_vs_extrapolation} and \ref{sec:parameter_sweep_2pt0}}
\label{tab:f2pt0_50pct_narrowing_study}
\begin{center}
\begin{tabular}{*{4}{c}}
\hline
  $r$  &  FEM                      & ALA (23)  & ALA (34)                  \\
\hline
-1.675 & 0.086                     & 0.085     & 0.086                     \\
\hline
-1.700 & \cellcolor{green!25}0.087 & 0.086     & 0.086                     \\
\hline
-1.725 & \cellcolor{green!25}0.087 & 0.086     & \cellcolor{green!25}0.087                     \\
\hline
-1.750 & 0.088                     & \cellcolor{green!25}0.087     & \cellcolor{green!25}0.087                     \\
\hline
-1.775 & 0.089                     & \cellcolor{green!25}0.087     & 0.088                     \\
\hline
-1.775 & Not Run                   & 0.088    &  0.088                     \\
\hline
\end{tabular}
\end{center}
\end{table}

\section{Discussion}
We have demonstrated that the novel ALA, which enables the neural network's training process to improve its own training set in a fully autonomous manner, successfully trains the neural network to accurately predict solutions to the Navier-Stokes equations everywhere in a parametric domain $\mathcal{R}$. Figures \ref{fig:l2_u}, \ref{fig:l2_p}, \ref{fig:l2_noslip_velocity} and \ref{fig:l2_inflow_velocity} strikingly demonstrate how the accuracy spreads across the parameteric domain $\mathcal{R}$ as more data are added, and in particular, to locations where no training data were provided. As an example application, in Section \ref{sec:parameter_sweep_1pt5} we demonstrated that the trained network can be used to make extremely accurate predictions about what tube shape will result in a particular physical property. It is clear that a neural network trained using these principles, encoding solutions to high-dimensional parametric descriptions of real-world physical problems, would be an extremely powerful tool in computational science, biomedical engineering, and even in medical applications where accurate real-time predictions are required.

\subsection{The Computational Efficiency of the Neural Network}
\label{sec:computational_cost}
Each of the two parameter sweeps described in Section \ref{sec:parameter_sweep} took 7.6 seconds; during this, we queried 81 points in $\mathcal{R}$, each at two points in space. Much of this time was taken up by loading the model and transferring it to the GPU; the scaling study shown in Table \ref{tab:scaling} demonstrates that the actual computation time per query is around 3 $\mu s$. Note however that we were not able to run 10,000,000 simultaneous queries, because the GPU ran out of memory (8.4 GB, due to a manual cap of 70\% of the 12 GB capacity, since the GPU also runs the test PC's display screens). Thus, beyond some hardware-dependent limit, queries would need to be batched. This is unlikely to present a real limitation.

For comparison, we observe that each Navier-Stokes FEM simulation used in this work took 40.5 seconds to run, thus a parameter sweep over 81 values of $r$ using FEM would have taken 54 minutes to compute, or over 400 times longer. This figure is provided for context only, as we did not attempt to optimise the speed of the FEM simulation, and this considers no optimal search strategy. 

\begin{table}[h!]
\caption{Time taken to query the trained neural network at two points in space, at various numbers of parameter points in $\mathcal{R}$. Each query consists of adjusting the domain shape (by adjusting $r$), and evaluating the pressure and velocity fields at two points in the domain, enabling the computation of a pressure difference with that particular domain shape.}
\label{tab:scaling}
\begin{center}
\begin{tabular}{*{7}{c}}
\hline
Parameter Queries & 40 & 80 & 160 & 10,000 & 100,000 & 1,000,000 \\
\hline
Run-Time (seconds) & 7.6 & 7.6 & 7.6 & 7.6 & 7.9 & 11.1\\
\hline
\end{tabular}
\end{center}
\end{table}

\subsection{Training Time}
Whilst Section \ref{sec:computational_cost} described the speed advantages to using a trained neural network, these must be considered in the context of the time taken to train the network initially. Approximately 22 hours were required to train the network for 23 ALA iterations, using an NVIDIA Titan Xp GPU.

However, network training occurs once, and then the network can be queried as and when new scientific questions arise, without need for re-training. Indeed, a key advantage of the ALA is that a network can be trained before the scientific question has been posed, provided that a sufficiently-encompassing $\mathcal{R}$ was chosen, and then questions can be rapidly answered as they arise. This is in contrast to the single-simulation FEM paradigm, which requires singular parameter choices, and thus generally a pre-posed scientific question.

It is worth noting that we used 100\% of all available velocity training data at each value of $\theta$ (i.e. velocity at all FEM nodes). This is unlikely to be necessary; it has been shown that sparser sampling of the data is sufficient for training in some cases \cite{MR19}.

\subsection{Alternative Training Strategies to the ALA}
We compared the ALA to the alternative strategy of random point selection, and we observed that - after the same number of training iterations - the ALA-trained network had a lower global loss $L_{NS}$, and the predictive accuracy of the ALA-trained network was better, compared to the random-trained network.

We did not compare ALA training against simply training on all points of a regular, coarse, fixed grid. We argue that using such a fixed training grid would be a poor strategy for a number of reasons. Foremost, in higher-dimensional parameter spaces, this would very quickly become extremely computationally expensive. Secondly, \textit{a priori} knowledge would be required to select the appropriate coarseness, and this coarseness would be limited by the fastest-changing behaviour anywhere in parameter space. Further, this fixed grid would waste computational effort in locations where the grid spacing was unnecessarily fine. The ALA addresses both of these problems: no \textit{a priori} information is required, and data will automatically be acquired more densely in regions of rapid parameter variation. Thus, the ALA has conceptual similarities to well-studied adaptive strategies in FEM \cite{CA12,CA13}.

\subsection{Interpolation vs. Extrapolation}
\label{sec:interpolation_vs_extrapolation}
In Section \ref{sec:parameter_sweep_1pt5}, we achieved very accurate predictions using the neural network trained for 23 ALA iterations. In Section \ref{sec:parameter_sweep_2pt0}, we found that at 23 ALA iterations the agreement with FEM data was good, but not quite as perfect. A likely reason for this lower accuracy is that in the latter case, the parameters of interest are at the edge of $\mathcal{R}$. Specifically, $r=-1.725$ and $f=2.0$ cause the network to extrapolate the training data in the $f$-direction (and interpolate in the $r$-direction); see Iteration 26 in Figure \ref{fig:l2_noslip_velocity}. Specifically, nothing to the right of this point in parameter space constrains it; in general, we should be wary of predictions near the edge of $\mathcal{R}$. Indeed, if we permit the ALA to run beyond the results presented in this paper, we observe that its longer-term behaviour is to focus on acquiring data at the boundaries of $\mathcal{R}$, rather than acquiring much additional data in the interior.

Noting that both $f$ and $r$ were interpolated in Section \ref{sec:parameter_sweep_1pt5}, the contrasting accuracy with Section \ref{sec:parameter_sweep_2pt0} supports the hypothesis that we must be careful at the boundary of the convex hull of training data.

\subsection{Number of Input Parameters, Data Aggregation and Perpetual Training}
In this work, our models were parametrised only by inflow rate and a single domain shape parameter, shown in Figure \ref{fig:neural_network}. In practical applications, hundreds of such parameters describing the model could be used, in addition to three-dimensional spatial parameters, and a time dimension. This could permit highly complex parametric geometries and boundary conditions to be encoded. Further work is required to determine the appropriate amount of training data required for such high-dimensional spaces, but it should be noted that whilst such parameter spaces might be high-dimensional, the region of interest in many cases will be quite small. 

Some well-defined parameter spaces are routinely and continually explored using FEM; patient-specific blood flow simulation is a good example. \cite{CA17,TM15,Alimohammadi2014}. Thus, workers are already generating data to well-characterise parameter space regions $\mathcal{R}$, and these data could be aggregated into neural networks as they are generated. Such a data source could augment that which is autonomously generated by the ALA.

It is possible to imagine the ALA running perpetually, without human interaction, so that the latest and best version of the network could be consulted whenever a simulation question arises. In the case of questions where the network does not have a good solution available for the required parameter set - as determined by the ALA loss function - FEM simulations could be run manually, and then that data provided to the ALA as an augmentation.

\subsection{Data Storage Cost}
In addition to the benefits of using a neural network to near-instantaneously compute PDE solution fields, when compared to simply storing and querying finite element results, the advantages with regard to data storage are also substantial. Without any attempts at optimisation beyond creating gzip-compressed tarballs of each, the trained neural network requires 1.9 MB of storage space. This contrasts to 157 MB for storing finite element solution fields at all 13$\times$13 points of $\mathcal{G}$ - or approximately a 99\% reduction in the required storage space. Even if we only stored four of the 13$\times$13 FEM solutions - which would be a poor FEM characterisation of $\mathcal{R}$ - this would still require twice as much storage space as the neural network. This would only become more substantial for three-dimensional simulations or higher-dimensional parameter spaces. Because solution data to three-dimensional problems can run to many hundreds of gigabytes, the potential importance of this data compression aspect of using neural networks in physics-based problems should not be underestimated.

\subsection{Pressure Data, Network Training and Ease of Implementation Validation}
During training, we used FEM solution data, which comprises both pressure and velocity solution fields. We trained only on the velocity data, with the exception of a single pressure node which was used to avoid the network adding arbitrary constants to its prediction the pressure field, which it otherwise may have done due to the fact that $p$ appears only as a gradient in the loss function (Equation \ref{eqn:lossfunction_navier_stokes}). This reference pressure was enforced in the loss function by Equation \ref{eqn:lossfunction_pressure}.

We could have trained using the full FEM pressure field, in addition to the velocity data. However, ignoring the pressure data provides a stronger validation of the method. Firstly, it shows the method is applicable if the data source contains only velocity data. Secondly, it provides additional validation of the input data: because Equation \ref{eqn:lossfunction_navier_stokes} is implemented as a single line of code, it is very easy to identify implementation bugs. This is in contrast to FEM solvers, where the implementation is scattered throughout the codebase, and is thus far more error-prone and difficult to verify. This is not an idle point; during initial development, the FEM pressure fields disagreed with the neural network's predictions. The correctness of the neural network's Navier-Stokes expression was instantly verifiable, so this led us to suspect the FEM solver. We found that the FEM solver was not outputting $p$, but rather, $p/\rho$. It is very unlikely that we would have identified this problem if we were using the pressure data during training.

Finally, the presence of the Navier-Stokes loss term in Equation \ref{eqn:lossfunction_navier_stokes} is critical for the ALA, since we must evaluate the pressure and velocity predictions, in terms of how well they \textit{together} satisfy Navier-Stokes, at locations of $\mathcal{G}$ where no training data are available.

\section{Conclusions}
We have presented a novel method of training a physics-informed neural network to predict Navier-Stokes pressure and velocity fields everywhere in a parametric domain, where the parameters represent the domain shape and a fluid boundary condition. The key contribution is the demonstration that this can be done with minimal training data, selected and automatically generated during training by our novel active learning algorithm (ALA). This required the coupling of a mesh generator and a numerical solver for the Navier-Stokes equations into the neural network training process. We demonstrated that the network can be used to perform extremely fast (3 $\mu s$ per parameter-point) and accurate parameter sweeps when searching for parameters which give particular physical properties. Thus the neural network provides a powerful method of solving inverse problems. In addition, its ability to aggregate many solutions which are currently routinely performed over a parameter space, and its ability to store these solutions efficiently (in our case, we observed a $\approx 99\%$ reduction in hard drive space requirements relative to compressed FEM data) and query them rapidly mean that such methods may have a role to play in computational fluid dynamics, and in computational physics in general.

\section{Acknowledgements}
We acknowledge the United Kingdom Department of Health via the National Institute for Health Research (NIHR) comprehensive Biomedical Research Centre award to Guy's \& St Thomas' NHS Foundation Trust in partnership with King's College London and King's College Hospital NHS Foundation Trust, the Wellcome/EPSRC Centre for Medical Engineering [WT 203148/Z/16/Z], and funding from the Wellcome Trust via King's College London. The basic code in this work was built on open source software \cite{raissi2017physicsI,raissi2017physicsII}. We gratefully acknowledge the support of NVIDIA Corporation with the donation of the Titan Xp used for this research.
\bibliographystyle{abbrv}
\bibliography{active_learning_paper}

\end{document}